\documentclass[aps,pre,superscriptaddress,floatfix,twocolumn,showpacs]{revtex4-1}

\usepackage{amsmath}
\usepackage{amsfonts}
\usepackage{amssymb}
\usepackage{graphicx}
\usepackage{bm}
\usepackage{color}
\usepackage{hyperref}
\usepackage{cases}
\usepackage{ulem}
\usepackage{multirow}
\usepackage{braket}

\usepackage{bm}

\def\stacksymbols #1#2#3#4{\def\theguybelow{#2}
	\def\verticalposition{\lower#3pt}
	\def\spacingwithinsymbol{\baselineskip0pt\lineskip#4pt}
	\mathrel{\mathpalette\intermediary#1}}
\def\intermediary#1#2{\verticalposition\vbox{\spacingwithinsymbol
		\everycr={}\tabskip0pt
		\halign{$\mathsurround0pt#1\hfil##\hfil$\crcr#2\crcr
			\theguybelow\crcr}}}

\begin{document}

\preprint{APS/123-QED}

\title{Discretized optical dynamics in one-dimensionally synthetic photonic lattice}

\author{Zengrun Wen}
 \author{Kaile Wang}
 \author{Baole Lu}
 \email{lubaole1123@163.com}
 \affiliation{State Key Laboratory of Photoelectric Technology and Functional Materials, International Collaborative Center on Photoelectric Technology and Nano Functional Materials, Institute of Photonics and Photon-technology, Northwest University, Shaanxi, Xi'an 710069, China}
 \affiliation{Shaanxi Engineering Technology Research Center for Solid State Lasers and Application, Provincial Key Laboratory of Photo-electronic Technology, Northwest University, Shaanxi, Xi'an 710079, China}
 \author{Xinyuan Qi}
 \affiliation{School of Physics, Northwest University, Xi'an 710069, China}
 \author{Haowei Chen}
\author{Jintao Bai}%
 \email{baijt@nwu.edu.cn}
\affiliation{State Key Laboratory of Photoelectric Technology and Functional Materials, International Collaborative Center on Photoelectric Technology and Nano Functional Materials, Institute of Photonics and Photon-technology, Northwest University, Shaanxi, Xi'an 710069, China}
\affiliation{Shaanxi Engineering Technology Research Center for Solid State Lasers and Application, Provincial Key Laboratory of Photo-electronic Technology, Northwest University, Shaanxi, Xi'an 710079, China}


\date{\today}

\begin{abstract}
Synthetic photonic lattice with temporally controlled potentials is a versatile platform for realizing wave dynamics associated with physical areas of optics and quantum physics. Here, discrete optics in one-dimensionally synthetic photonic lattice is investigated systematically, in which the light behavior is highly similar to those in evanescently coupled one-dimensional discrete waveguides. Such a synthetic dimension is constructed with position-dependent periodic effective gauge fields based on Aharonov-Bohm effect arising from the phase accumulations of the fiber loops. By tuning the phase accumulations and coupling coefficient of the coupler,  the band translation and gap property can be modulated which further results in the impulse and tailored Gaussian wave packet responses as well as Talbot recurrences. In addition, Bloch oscillations and Anderson localization can also be obtained when the phase accumulations are linearly changed and weakly modulated in random, respectively. The periodic effective gauge fields configuration in our protocol enables SPL to be a research platform for one-dimensional dynamically modulated elements or even non-Hermitian waveguides.

\end{abstract}

\maketitle

\section{Introduction}
 Light propagation behaviors in discrete waveguides or photonic lattices have drawn considerable attention in last few decades. Generally, the arrangements of artificial periodic potentials exhibited remarkable advantage in controlling the aspects of light with their continuous counterparts due to the judiciously engineered diffraction~\cite{christodoulidesnature03}. The interaction of periodicity and abundant modulation fashion enables the coupled waveguides to support various light propagating phenomena. Among them, discrete diffraction, anomolous refraction and discrete Talbot effect are the fundamental behaviors of a linear homogenous waveguide array~\cite{pertschprl02,iwanowprl05}.  In the modulated framework, Bloch oscillation and Anderson localization can be separately achieved in linearly modulated potentials and disordered waveguides, providing a way to comprehend modulation-driven diffraction management of light~\cite{preshelol98,pertschprl99,lahiniprl08,segevnp13}. Experimental implementation of these dielectric waveguides arrays needs sophisticated fs-laser lithography on the nanometer or micron scale~\cite{chenlpr14}. Recently, the study of coupled waveguides is extended to non-Hermitian or ${\cal PT}$-symmetric photonic lattices~\cite{eiganainyol07,makrisprl08,remezaniprl12,longhiprl09,kartashovlpr16}. The much complex experimental manipulation in balancing the gain and loss hinders the progress of light control in the associated field~\cite{rueternp10}. 
 
An increasing research area on synthetic photonic systems such as dynamically modulated resonators, multidimensional synthetic lattices and synthetic photonic lattice (SPL) emerged for reducing the difficulty or breaking through the limitation of the experiments with respect to the corresponding simulated quantum or optical systems~\cite{yuanoptica,duttscience,yuanprl,maczewskynp,regensburgerprl}.  Utilizing the time-multiplexing technique in one coupler connected two coupled fiber loops, the SPL is constructed and provides a high degree of adjustability and easy-manipulated experimental platform in macroscope for studying quantum walks and optical dynamics in periodic discretized structures~\cite{regensburgernature}. Various parameters including light intensities and phases (corresponding to the imaginary and real part of potentials), as well as the coupling coefficient of the coupler can be modulated in order to control the flow of light flexibility~\cite{miripra,pankovoe}. By engineering these parameters, a number of special phenomena have been investigated in SPLs, diametric drive acceleration~\cite{wimmernp13}, time mirror~\cite{wimmersr18} and kapiza light~\cite{munizol19}. Naturally, the discretized optical behaviors associated with discrete arrays such as discrete diffraction, Talbot effect~\cite{wangpra}, Bloch oscillation~\cite{wimmersr15}, Anderson localization~\cite{vatniksr17}, soliton~\cite{wimmernc15} and other localized states with respect to defect, gauge fields and topological phase transition~\cite{regensburgerprl13,pankovsr19,bisianovpra19,weidemannscience} have been realized in the SPLs under passive or ${\cal PT}$-symmetric configurations.  However, the theoretical model in SPL, in which the light dynamics is highly similar to that in conventional one-dimensional photonic lattices, has not yet been established systematically.

 In this work, we construct a position-dependent periodic gauge fields in the SPL. The gauge field induces two phase accumulations with counter direction distribute alternatively in the transverse direction. By changing the intensity of phase accumulation and the coefficient of the coupler, the variation of band structure, bandgap and the similar propagation condition of impulse in the two fiber loops are investigated. Interestingly, analogues of the discrete diffration, anomolous refraction and Talbot recurrence are achieved for the tailored Gaussian beam and periodic sources injection. Besides, Bloch oscillations and Anderson localization are numerically achieved when SPL are gradiently modulated and weakly disordered on the basis of periodic gauge fields, respectively, further proving that the knowledge in conventional photonic lattice should be meaningful in the light dynamic in SPL, too. 
 
 \section{The model and band structure of photonic mesh lattice with periodic gauge fields}
 \begin{figure}[b]
 	\includegraphics{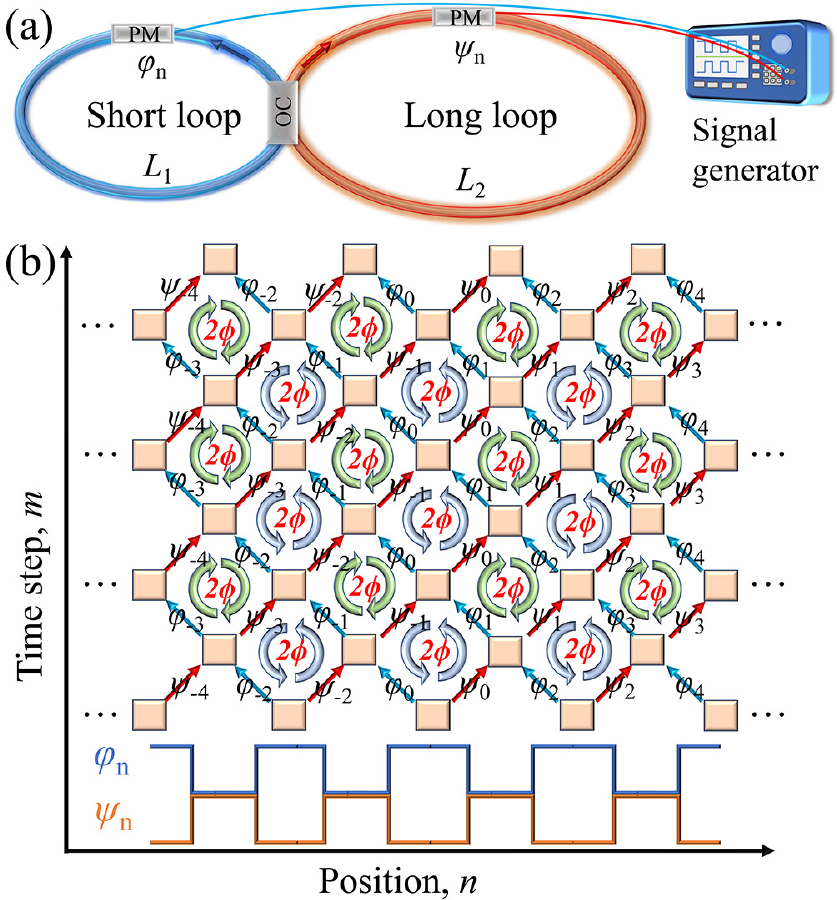}
 	\caption{\label{fig1} (a) Schematic of the SPL realized by two mutually coupled fiber loops. (b) Equivalent SPL corresponding to the coupled fiber loops. The phase configuration in the bottom is controlled by the signal generator.}
 \end{figure}

 One of the most typetical SPLs is realized by two appropriately designed fiber loops connected by a  directional fiber coupler. As shown in Fig. \ref{fig1}(a), the two fiber loops have a imbalanced path length $\Delta L$ which gives raise to discretized arrival times $2\Delta T=\Delta L/c_F$, where $c_F$ represents the light speed in the fiber. When an optical signal is injected in the setup, it will be separated into two signals after the coupler, then the two light propagate along the short (blue) and long (pink) fiber ring, respectively. The  pulse travels along long (short) ring corresponds to +1 (-1) of the position $n$ in the transverse direction of SPL. The interference of the two pulses happens at the coupler, and the count/time of interference is described by $m$ in the longitudinal direction of SPL [see Fig. \ref{fig1}(b)]. According to the above principle, the setup is translated to time-multiplexed scheme, namely, the SPL with infinite extension along transverse and longitudinal direction. In our work, two phase modulators driven by a dual channel signal generator are respectively adopted in the long and short fiber rings to alter the refractive index of the SPL, which are shown in Fig. \ref{fig1}(a). The amplitude of light in SPL is governed by the following iteration equations:
 \begin{eqnarray}
 \begin{split}
 u_n^{m+1}=e^{i \varphi_n}[\cos(\theta)u_{n+1}^m+i\sin(\theta)v_{n+1}^m], \\
 v_n^{m+1}=e^{i \psi_n}[\cos(\theta)v_{n-1}^m+i\sin(\theta)u_{n-1}^m],
 \end{split}\label{eq1}
 \end{eqnarray}
where $u_n^m$ and $v_n^m$ are the complex amplitudes of pulses circulating in the short and long loops, respectively. The position-relevant phases $\varphi_{n}$ and $\psi_{n}$ are provided by the phase modulators in the short and long loops, respectively. $\theta$ describes the coupling ratio of the fiber splitter. The loss of this setup is not taken into consideration because it can be compensated by inline semiconductor amplifiers in practical operation~\cite{wimmernc15}.

To establish a gauge filed in the SPL, the sign of phases ($\varphi_{n}$, $\psi_{n}$) in the two fiber loops ought to be opposite.  Besides, the single gauge field induces phase circumfluence in the corresponding site and the adjacent sites~\cite{pankovsr19}. For avoiding the interruption of the neighboring phase circumfluence, the interval $P_n$ of the gauge fields is at least two position sites. In this work, we discuss the condition of $P_n=2$, the phase configuration is shown in the bottom of Fig.~\ref{fig1}(b). Correspondingly, the phase expression in the two loops reads
\begin{equation}
\varphi_n=\left\{
\begin{array}{lr}
\phi, & \mod(n,2)=0 \\
0, & \mbox{otherwise}
\end{array}
\right.
,\psi_n=-\varphi_n.\label{eq2}
\end{equation}
Here, $\psi_n\equiv-\varphi_n$ is fixed to reduce the complexity without loss of generality. The circumfluence intensity is $\varphi_n+\psi_{n-1}-\varphi_{n+1}-\psi_{n}$, viz, $2\phi$ and $-2\phi$ in even and odd sites, respectively. As a result, the counter-clockwise and clockwise phase accumulations distribute alternatively along the position direction which are intuitively depicted in Fig.~\ref{fig1}(b). Two types of phase accumulation indicate there are two kinds of waveguide.

\begin{figure}[b]
	\includegraphics{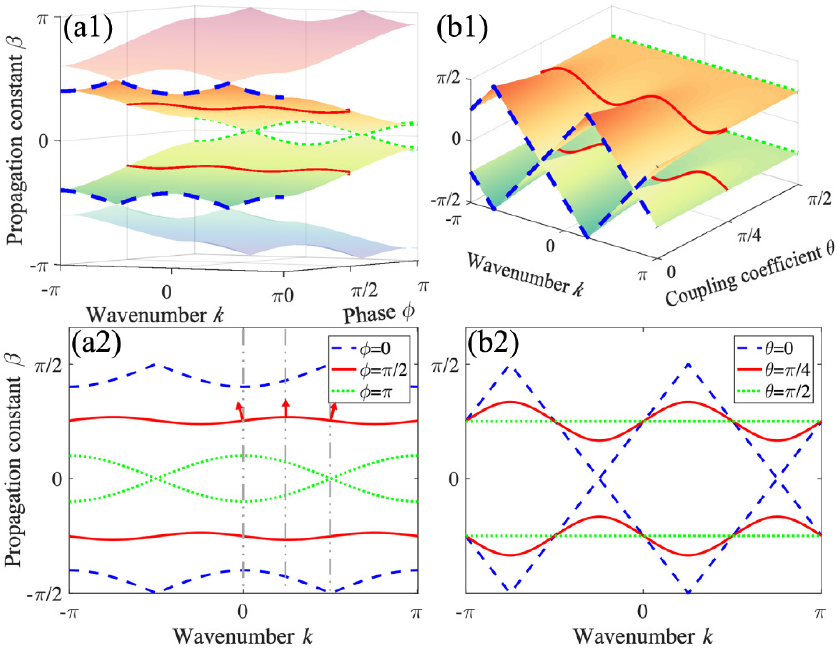}
	\caption{\label{dispersion-relation} Band structures of the SPL in the present of periodic gauge fields versus (a1)-(a2) phase circumfluence $\phi$ and (b1)-(b2) coupling coefficient $\theta$.}
\end{figure}
The dynamic properties of a SPL system can be extrapolated by its band structure (dispersion relation) in wavenumver space. To obtain the band structure of the SPL, a Floquet-Bloch discrete plane-wave ansatz is used as the solution of Eqs. (\ref{eq1}), which is
\begin{equation}
\left(
\begin{array}{c}
u_n^m\\v_n^m
\end{array}
\right)=
\left(
\begin{array}{c}
U\\V
\end{array}
\right)\exp(i k n+i \beta m),\\
\label{eq3}
\end{equation}
where $\beta$ is the propagation constant and $k$ designates the transverse wavenumber. Substituting Eq. (\ref{eq3}) into Eq. (\ref{eq1}) leads to the eigenvector problem
\begin{equation}
H(\theta,\phi)\left(
\begin{array}{c}
U\\V
\end{array}
\right)=\lambda\left(
\begin{array}{c}
U\\V
\end{array}
\right),\label{eq4}
\end{equation}
in which the eigenvalue $\lambda=e^{i\beta}$ is an intrinsic characteristic of SPL~\cite{regensburgerprl13}. The $H(\theta,\phi)$ describes the Hamiltonian of the SPLs, which is 
\begin{small}
\begin{equation}
\left(
\begin{array}{cccc}
0 & 0 & e^{i(k+\phi)}\cos(\theta) & ie^{(k+\phi)}\sin(\theta)\\
0 & 0 & ie^{-(k+\phi)}\sin(\theta) & e^{-i(k+\phi)}\cos(\theta)\\
e^{ik}\cos(\theta) & ie^{ik}\sin(\theta) & 0 & 0\\
ie^{-ik}\sin(\theta) & e^{-ik}\cos(\theta) & 0 & 0
\end{array}
\right).
\label{eq5}
\end{equation}
\end{small}

When the phase configuration follows Eq. (\ref{eq2}). The dispersion relation is calculated by solving Eqs. (\ref{eq4}) and (\ref{eq5}), which reads
\begin{equation}
\beta_\pm=\pm\frac{1}{2}\arccos[\cos^2(\theta)\cos(2k+\phi)-\cos(\phi)\sin^2(\theta)].\label{eq6}
\end{equation}
The above equation manifests $\beta$ is a periodic function with period of $\pi$. The band structure variates two periods with regard to the wavenumber $k$ from $-\pi$ to $\pi$, which is depicted in Fig. \ref{dispersion-relation}. In general, for arbitrary values of $\theta$ and $\phi$, the dispersion relationship is valid. However, the band structure in the domain of $0<\phi<\pi$ and $0<\theta<\pi/2$ reflects the complete principles in the whole domain.
Figures \ref{dispersion-relation}(a1) and (a2) depicts the band structure in the first Brillouin zone ($-\pi/2<\beta<\pi/2$ and $-\pi<k<\pi$) versus phase circumfluence $\phi$ when $\theta$ is fixed at $0.42\pi$. There are two bands distribute symmetrically with respect to $\beta=0$, and both degenerate at $\phi=0$ (occurring at $\beta=\pm\pi/2$) and $\phi=\pi$ (occurring at $\beta=0$). When $0<\phi<\pi$, the bandgap between the two symmetric bands arise and reach its maximum as $\phi=\pi/2$. According to the term $\cos^2(\theta)\cos(2k+\phi)$ in Eq. (\ref{eq6}), one can see the value of $\phi$ also induce the linear translation of the band structure with respect to wavenumber. The transition makes the maximum value of $\beta$ transit from $k=\pm\pi/2$ to $k=0$ when $\phi$ increases from $0$ to $\pi$. Besides, the maximum bandgap exists at $\phi=\pi/2$, corresponding to non-symmetric band structure induced by band translation. The variation of the band structure versus $\theta$ is described in Figs. \ref{dispersion-relation}(b1) and (b2) when $\phi$ is fixed at $\pi/2$. With the value of $\theta$ increasing from $0$ to $\pi/2$, the two bands begin to saparate, then tend to gentle and finally flatten at $\theta=\pi/2$. Meanwhile, the bandgap enlarged.  

When the periodic gauge fields are introduced in SPL, there are always two energy bands in first Brillouin zone with variable $\phi$ and $\theta$, which is similar to that of coupled waveguides~\cite{}. The variation of the band structure in our work shows the SPL can be used as a platform for controlling light dynamics.

\section{Optical dynamics in SPL with periodic gauge fields}
In this section, we discuss optical dynamics under three different initial excitations comprising of single-site, tailored Gaussian beam and periodic point sources. Since the impulse can excite the entire band, which is utilized to study the propagation properties of the SPL with different $\theta$ and $\phi$. Tailored wave packet dynamics reflect the group velocity and dispersion modulation of the SPL. Talbot effect of the system is achieved with periodic source input. 
\subsection{Point source responses}
\begin{figure}[b]
	\includegraphics{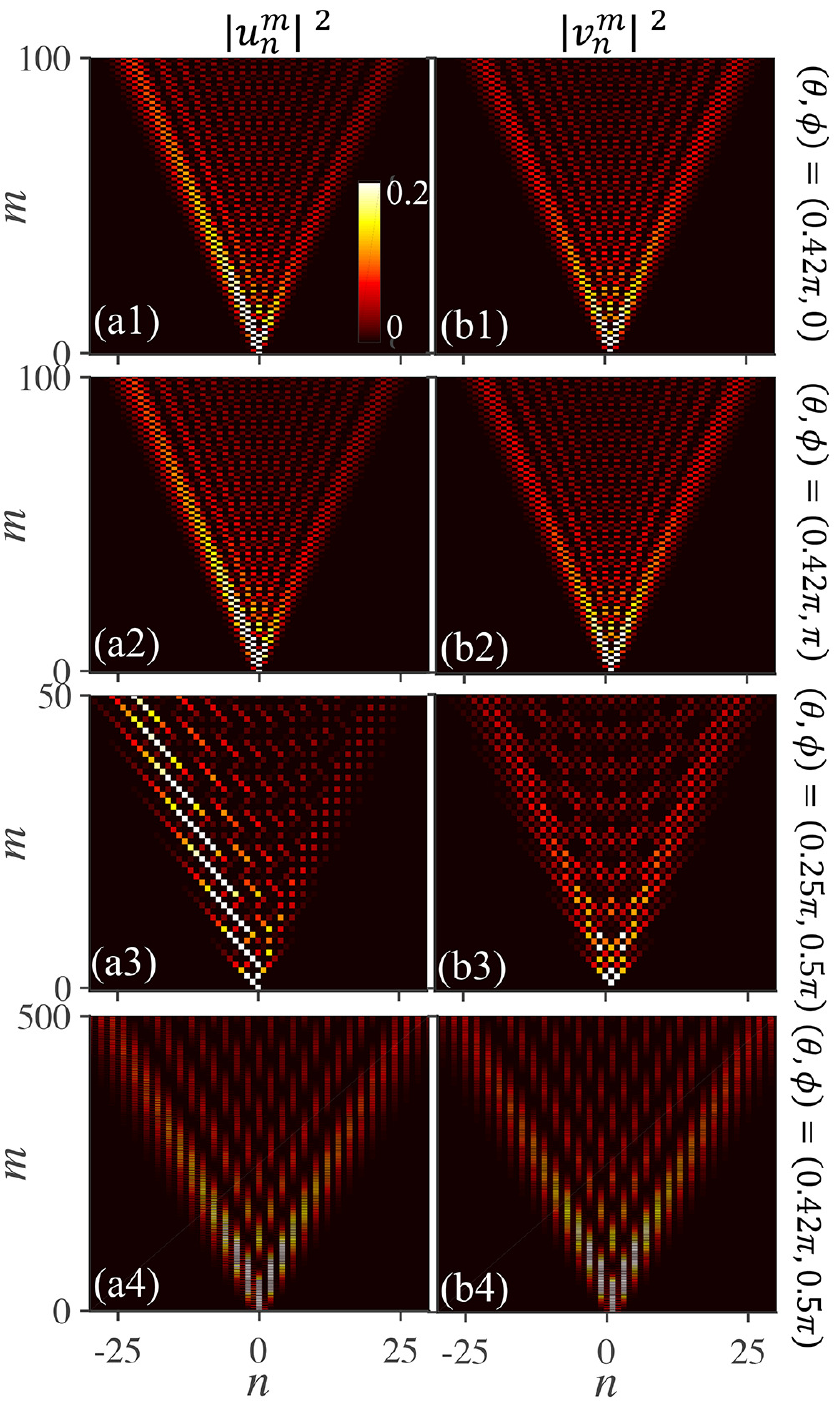}
	\caption{\label{discrete-diffraction} Impulse responses in the SPL with different $\theta$ and $\phi$. The left and right columns are light trajectories in short loop $|u_n^m|^2$ and long loop $|v_n^m|^2$.}
\end{figure}
Light responses in the two fiber loops are saparately described by two propagation diagrams $|u_n^m|^2$ and $|v_n^m|^2$. Thus, the relation of optical dynamics in the two fiber loops is explored firstly to reduce the discussion complexity. To solve the problem, Eq. (\ref{eq1}) is further deduced as
\begin{eqnarray}
\begin{split}
u_n^{m+2}&=e^{i \varphi_{n}}[\cos(\theta)u_{n+1}^{m+1}\\
&+e^{i \psi_{n+1}-i\varphi_{n-1}}\cos(\theta)u_{n-1}^{m+1}-e^{i\psi_{n+1}}u_n^m], \\
v_n^{m+2}&=e^{i \psi_{n}}[e^{-i \psi_{n+1}+i\varphi_{n-1}}\cos(\theta)v_{n+1}^{m+1}\\
&+\cos(\theta)v_{n-1}^{m+1}-e^{i\varphi_{n-1}}v_n^m]. \label{eq7}
\end{split}
\end{eqnarray}
In the equations, optical intensities in the two fiber loops are separately expressed by the iteration forms. Obviously, once similar expression between $|u_n^{m+2}|^2$ and $|v_n^{m+2}|^2$ is guaranteed, there is $\varphi_{n-1}-\psi_{n+1}=$0 or $j\pi$ ($j$ is an integer). As a result, the phase accumulation $2\phi$ in Eq. (\ref{eq2}) is limited to 0 or $j\pi$ and the transport behaviors in two fiber loops ought to be similar. 

 As a point source is injected at position site $n=0$, the light dynamical propagations in the SPL are shown in Fig. \ref{discrete-diffraction}. Clearly, the similar propagation trajectories are displayed between the left column $|u_n^m|^2$ and right column $|v_n^m|^2$, only with one position site transition which is caused by initial condition delay of $v_n^m$ comparing with $u_n^m$. The top two rows of Fig. \ref{discrete-diffraction} depict the ballistic transport corresponding to phase circumfluence of 0 and $2\pi$, respectively, all having zero bandgap in band structure. Thus, the light transport continuously to their nearest-neighboring sites. When $\theta=0.25\pi$ and $\phi=0.5\pi$ are set, continuous spread between adjacent position sites still exists. This is attributed to the low localized capacity of the gauge fields with little $\theta$~\cite{pankovsr19}. For a larger $\theta$ of $0.42\pi$, gauge fields induced strong localization of sites arises and light hops between the position sites possessing identical phase circumfluence and nearly no light energy exists in odd position sites [see Figs.~\ref{discrete-diffraction}(a4) and (b4)]. Comparing the four cases in Fig.~\ref{discrete-diffraction}, the bottom case is closer to the single-site excitation case in one-dimensional discrete waveguide with evanescent field coupling.

 In the rest of the work, the special case of $\theta=0.42\pi$ and $\phi=0.5\pi$ are settled. Besides, only the optical dynamics of the fiber loop with initial input are considered for simplicity. 
 
\subsection{Tailored Gaussian wave packet responses}
\begin{figure}[b]
	\includegraphics{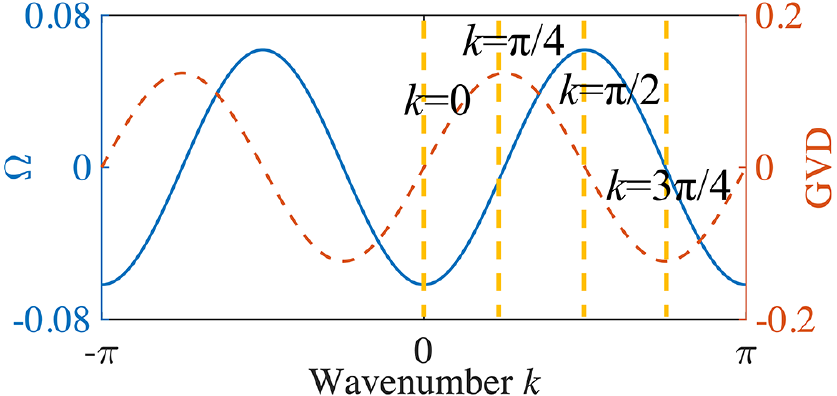}
	\caption{\label{gvd} Group velocity and GVD of the SPL versus $k$.}
\end{figure}
\begin{figure}[b]
	\includegraphics{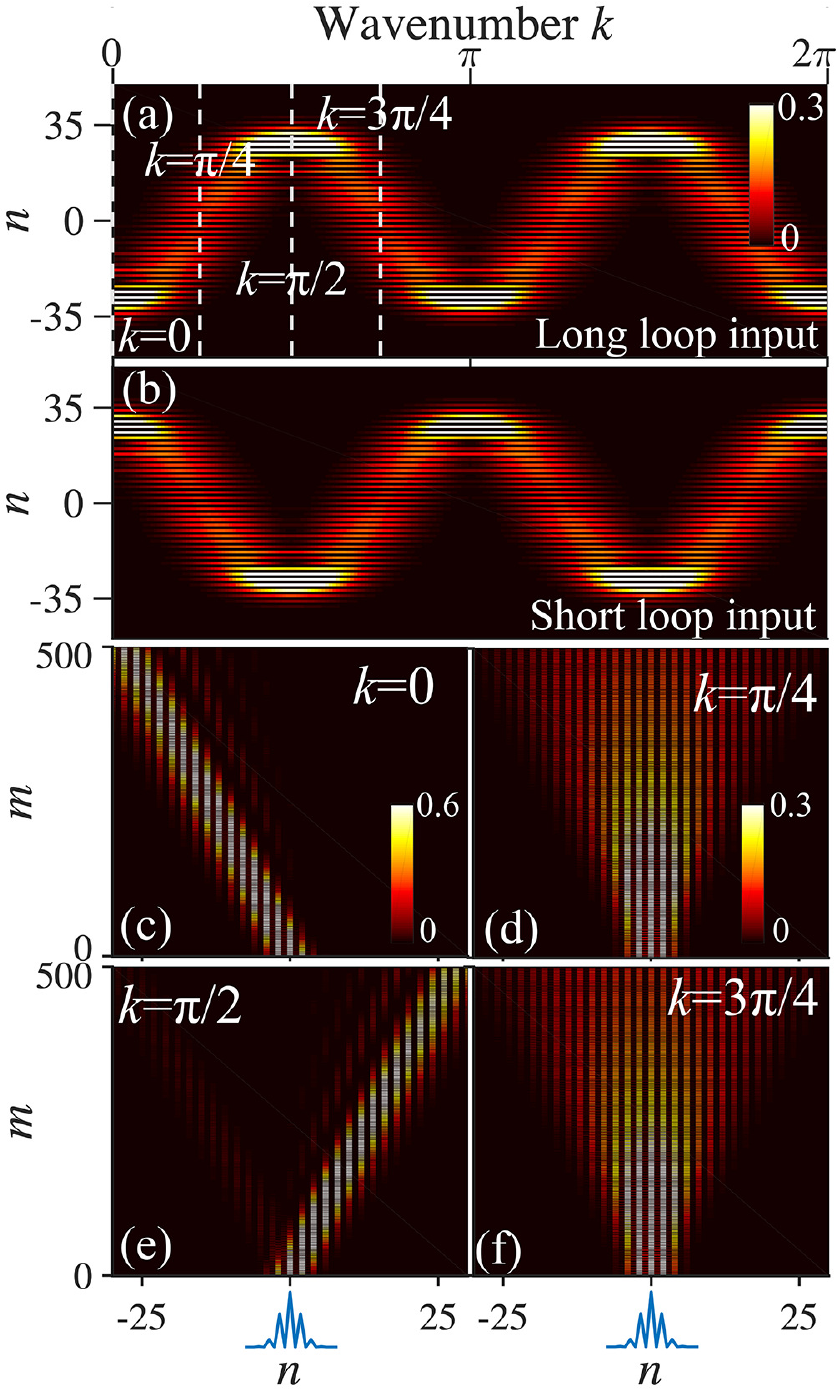}
	\caption{\label{gaussian-input} Taiored Gaussian wave packet responses in the SPL. (a) and (b) are output correspinding to long and short loops excitation as a function of wavenumber $k$, respectively. (c)-(f) are light dynamics in the short loops with four chosen wavenumbers $k$.}
\end{figure}

When the SPL is excited by a broad wave packet, the group velocity ($\Omega$) and group velocity dispersion (GVD) of the envelope associated with the first and second differential terms in Taylor expansion of dispersion relation~\cite{pertschprl02}, which are 
\begin{equation}
\Omega=\frac{d\beta}{dk_{|k_0}}=-\frac{\cos^2\theta\cos(2k)}{\sqrt{1-\cos^4\theta\sin^2(2k)}}
\label{eq8}
\end{equation}
and
\begin{equation}
GVD=\frac{d^2\beta}{dk^2_{|k_0}}=-\frac{2\cos^2\theta(\cos^4\theta-1)\sin(2k)}{{[1-\cos^4\theta\sin^2(2k)]^{3/2}}},
\label{eq9}
\end{equation}
respectively. Figure~\ref{gvd} plots the group velocity and GVD as a function of wavenumber when $\theta=0.42\pi$. Under this circumstance, the values of denominators in Eqs. (\ref{eq8}) and (\ref{eq9}) are close to 1. Thus in the approximate, the group velocity and GVD are cosine and sine functions, respectively. 
For several typical wavenumbers, the group velocity and GVD are marked in Fig.~\ref{gvd}.  As a normal incident beam ($k=0$) is injected, the group velocity is maximum in the negative direction, whereas the GVD is zero. This indicates the beam will slant leftward directionally without diffraction. Inversely, non-diffraction transmission of the beam with rightward tilt is formed if the wavenumber $k$ is $\pi/2$. The maximum positive and negative GVD corresponds to $k=\pi/4$ and $k=3\pi/4$, respectively, where the group velocities of the beam are zero. As a result, the beam will spread perpendicularly to the incident plane and experiences diffraction. 

Generally, a Gaussian wavep acket excitation was used to investigate the anomalous refraction and diffraction properties of a discrete array~\cite{pertschprl02}. In this part, a specially customized tailored Gaussian wave packet is considered for studying the diffraction effects of the SPL, which is
\begin{equation}
u_0(n)=\left\{
\begin{array}{lr}
e^{-(n/\Delta)^2}e^{ikn}, & \mod(n,2)=0\\
0, & \mod(n,2)=1
\end{array},
\right.
\label{eq10}
\end{equation}
where $2\Delta$ represents the width of pulse envelope. The amplitude of the tailored Gaussian wave packet is zero when the position site $n$ is odd where clockwise phase accumulations locate, all being treated as coupling areas in this work simultaneously. Figure~\ref{gaussian-input} shows the optical dynamics of this tailored Gaussian beam with pulse width $2\Delta=8$ in the SPL. From Fig.~\ref{gaussian-input}(a), when the beam is initially excited in the long loop, the output pulse profiles are highly dependent on wavenumber $k$ and both the output position sites and period versus $k$ are coincident with the group velocity depicted in Fig.~\ref{gvd}. Besides, there is a maximum reachable position site and the pulse width varies with the input angle. For the case of short loop injection, the propagation trajectories are mirror symmetry with those of the long loop input, corresponding to the lower band in Fig.~\ref{dispersion-relation}. In order to verify the discussion concerning diffraction and GVD, tailored Gaussian beam responses with four chosen input angle $k$ are numerically simulated and displayed in Figs.~\ref{gaussian-input}(c-f). For the cases of $k=0$ and $k=\pi/2$, the transmission trajectories of the beam tilt to the left and right without dispersion. The tilt angles are about -0.062 rad and 0.062 rad, respectively. While according to Figs.~\ref{gaussian-input}(d) and (f), the beam propagates forward and broadens rapidly. Evidently, the optical dynamics of the tailored Gaussian beam accord with the analysis of dispersion and GVD. More importantly, the propagation traces are also closely similar to those in discretized waveguides~\cite{pertschprl02}. 

\subsection{Talbot effect}
\begin{figure}[b]
	\includegraphics{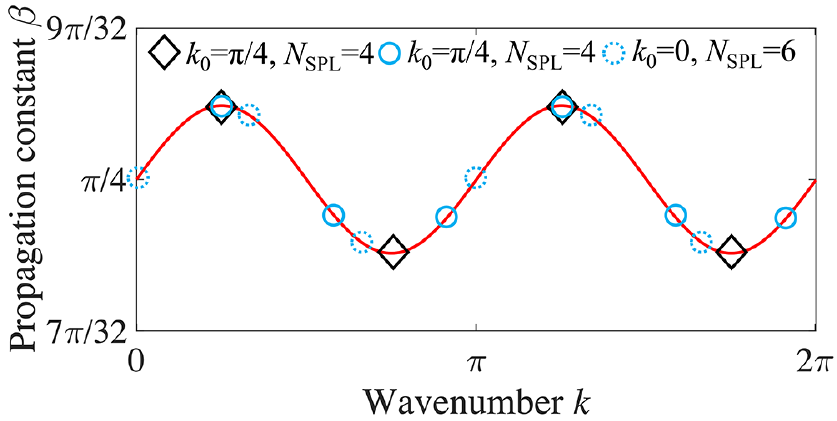}
	\caption{\label{discretek} Discrete wavenumbers and propagation constants limited by periodic sources. $k_0$ is the initial angle of input source.}.
\end{figure}
\begin{figure}[b]
	\includegraphics{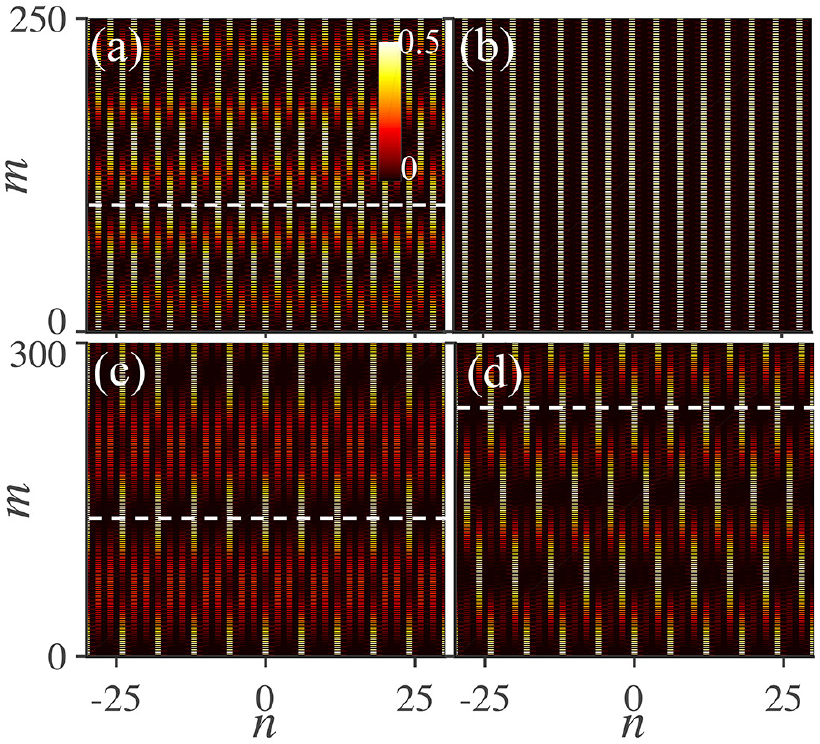}
	\caption{\label{talbot} Talbot revivals in the SPL for  (a) $e^{i\pi/4}$\{1 0 0 0 1 0 0 0 1 ...\}; (b) $e^{i\pi/4}$\{1 0 0 0 -1 0 0 0 1 ...\} ; (c) $e^{i\pi/4}$\{1 0 0 0 0 0 1 0 0 0 0 0 1 ... \}; (d) \{1 0 0 0 0 0 1 0 0 0 0 0 1 ...\} periodic source excitations.}.
\end{figure}
Above subsections show the similarity of our SPL and one-dimensional waveguide array. In this part, we will study the connection of the two system with respect to Talbot recurrence.

In a discrete waveguide array, the Talbot revivals occurs only for the periods $N$ of initial source in the range $\{1,2,3,4,6\}$. Also, for inphase input, the Talbot recurrence distance is the largest period of $\xi_T=2\pi/(\lambda_i-\lambda_j)$, where $\lambda_i$ and $\lambda_j$ ($i$, $j\in\{0,1,...,N-1\}$) stand for the discrete eigenvalue limited by periodic source limited wavenumber $k$~\cite{iwanowprl05}.

Considering the plane-wave ansatz of Eq.~\ref{eq3}, the periodic input condition limits $v_n^m=v_{n+N_{\mbox{SPL}}}^m$, giving a discrete wavenumber $k_j=2j\pi/N_{\mbox{SPL}}$, , $j\in\{0,1,...,N-1\}$. In the equation, $N_{\mbox{SPL}}$ describe the period of input source. Here, we investigate the discrete Talbot effects by means of the example of $N_{\mbox{SPL}}=4$ and 6, corresponding to $N=2$ and 3 for discrete waveguides, respectively. According to the band structure analysis of the SPL, the translation of the band structure is $\pi/4$ when the phase accumulation $2\phi=\pi$. Therefore, in order to compensate for the translation, the angle of the injected sources are modulated as $k_0=\pi/4$ firstly. Figure~\ref{discretek} plots the discrete wavenumbers $k_j$ and propagation constants $\beta_j$ with different $k_0$ and $N_{\mbox{SPL}}$. As the equivalent between discrete waveguides and the SPL, the Talbot distance for inphase source is the maximum period of 
\begin{equation}
m_T=2\pi/(\beta_i-\beta_j).\label{eq11}
\end{equation}
According to the equation, for $k_0=\pi/4$, $N_{\mbox{SPL}}=4$ and $k_0=\pi/4$, $N_{\mbox{SPL}}=6$, the Talbot distance are $2\pi/(\beta|_{k=\pi/4}-\beta|_{k=3\pi/4})\approx102$ and $2\pi/(\beta|_{k=\pi/4}-\beta|_{k=7\pi/12})\approx135$ when the inputs are inphase. The corresponding Talbot carpets are depicted in Figs.~\ref{talbot}(a) and (c), respectively. Every point source is symmetrically coupled to its neighbor sites with gauge fields due to the dispersion and reappears at Talbot distances. If the input source has different initial phase, a special case arises when the input pattern is $e^{i\pi/4}$\{1 0 0 0 -1 0 0 0 1 ...\} where there is a phase difference $\pi$ between the adjacent sources. As shown in Fig.~\ref{talbot}(b), the light propagates straightforwardly and the Talbot recurrence disappears. This kind of input is equal to the one in which the phase difference between nearest position sites is $\pi/4$ and all the discrete propagation constans are $\beta_j=\pi/4$. Therefore, there is no Talbot distance. When the initial angle is changed to $k_0=0$ (eg. input pattern of \{1 0 0 0 0 0 1 0 0 0 0 0 1 ...\}), $\beta_j$ shifts to the left by $\pi/4$, the Talbot distance is altered to $2\pi/(\beta|_{k=\pi/3}-\beta|_{k=0})\approx235$ [see Fig.~\ref{talbot}(d)]. The input sources are coupled to their left position sites with identical phase circumfluence due to the negative group velocity and reemerge after three times of coupling. All the numerically simulated Talbot distances are coincident with the calculation based on Eq.~\ref{eq11}. 

\section{Optical dynamics in the SPL with modulated Gauge phase accumulations}
In this section, we focus on the optical dynamics in the SPL when the phase accumulations are modulated in the position axis. Bloch oscillations and Anderson localizations can both be achieved as an linearly change and random modulation in the phase circumfluence are introduced, respectively. 
\subsection{Bloch oscillation}
\begin{figure}[b]
	\includegraphics{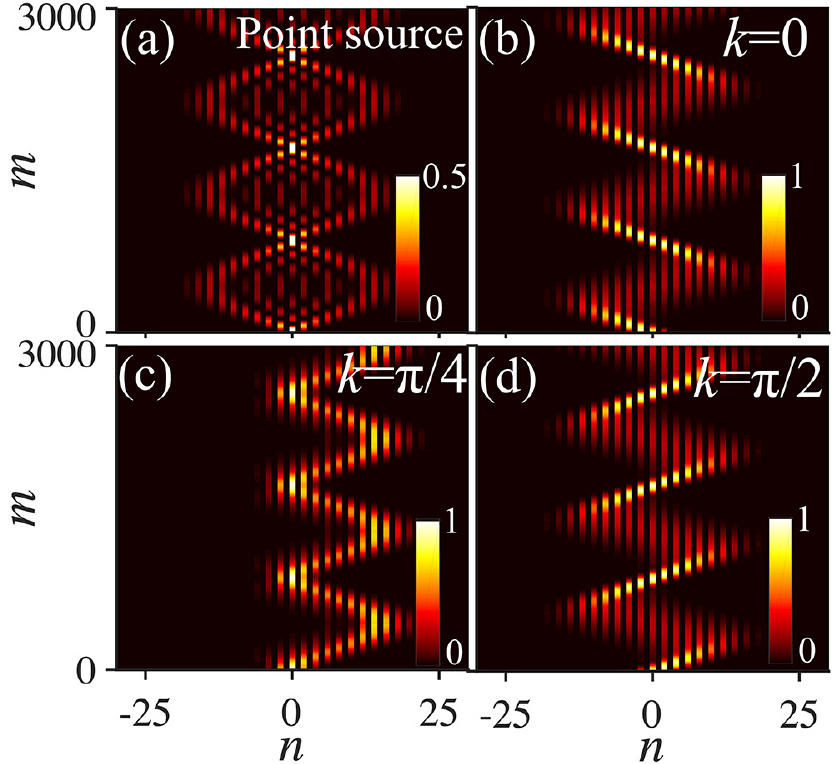}
	\caption{\label{bloch} Bloch oscillations in the SPL with linear modulated phase $\phi$. (a) point source excitation; (b-d) Gaussian wave packet with different $k$ as input.}
\end{figure}
\begin{figure}[b]
	\includegraphics{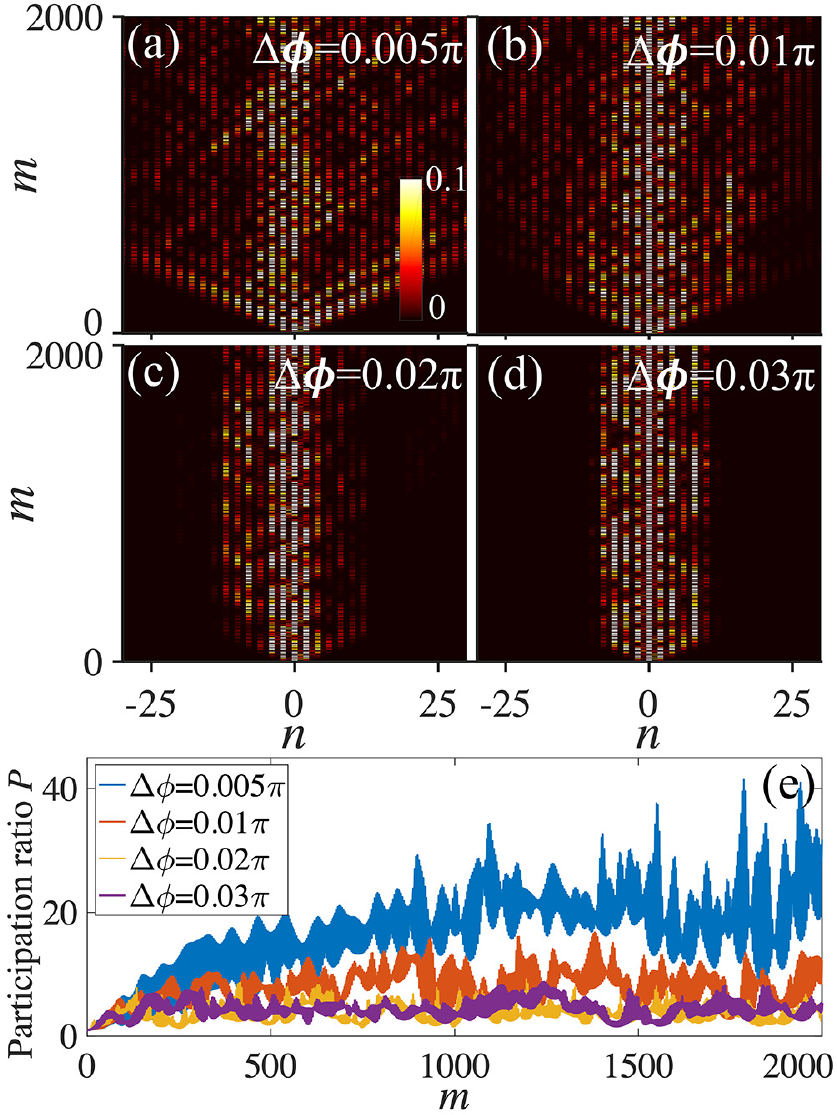}
	\caption{\label{anderson} Anderson localizations in the SPL with increasing phase disorders. (a-d) maximum phase disorders of $\Delta\phi$ of $0.005\pi$, $0.01\pi$, $0.02\pi$ and $0.03\pi$. (e) paticipation ratios corresponds to (a-d).}
\end{figure}
Once there is a linearly variated effective refractive index in a waveguide array, an occurrence of periodic oscillation termed Bloch oscillation was theoretically predicted and experimentally achieved~\cite{preshelol98,morandottiprl99}. The physical mechanism is increasing phase difference (wavevector of light) between adjacent waveguides and a lateral shift of the beam position caused by the linear index gradient factor.  In the SPL, the equivalent manipulation of effective refractive index is the intensities of the phase potentials. Unlike the proposed Bloch oscillations in the SPL with time step relevant gradient phases~\cite{wimmersr15}, the phase circumfluence in our SPL is linearly modulated in position direction on the base of periodic gauge fields, which is a direct analogy from a waveguide array. The phases in the two loops follows
\begin{equation}
\psi_{m,n}=\left\{
\begin{array}{lr}
\phi-n\alpha, & \mod(n,2)=0 \\
0, & \mbox{otherwise}
\end{array}
\right.
,\varphi_{n}=-\psi_{n},
\end{equation}
where $\alpha=\pi/400$ is the linear attenuation factor. For the neighboring gauge fields, the attenuation rate is $2\alpha$. According to Eq.~\ref{eq6}, the variation of phase $\phi=2\alpha$ equals to the translation of wavenumber $\Delta k=\alpha$. Therefore, the wavenumber or wavevector direction is position site dependent. The propagating direction of light is altered in the coupling process between different gauge fields. Once the light returns to the initial position sites and acquires the initial wavevector simultaneously, the optical trajectory repeat again. Finally, the periodic oscillation is formed. 

Numerically results of Bloch oscillations are shown in Fig.~\ref{bloch}. As a point excitation, bidirectional Bloch oscillation is the displayed [see Fig.~\ref{bloch}(a)]. The light expands to several position sites with gauge fields bidirectionally due to the discrete diffraction. Then the direction is reversed after the beam broadening to $n=\pm16$. At a oscillation period, both the point source and initial condition are reset. When a broad tailored Gaussian beam with different wavenumbers is injected, the single Bloch oscillations are obtained, as depicted in Figs.~\ref{bloch}(b-d), the revival period is independent with the initial wavenumbers obviously. For $k=0$ and $k=\pi/2$, the initial pulse direction tilts leftward and rightward, respectively [see Figs.~\ref{bloch}(b) and (d)], which are coincident with the their directions of group velocities. Besides, the lights are limited between position sites of $-16<n<16$. In the above conditions, the initial direction affected by weak attenuation factor is covered by strong diffraction. While for $k=\pi/4$ (zero group velocity), the beam propagates rightward initially and is limited in the minimum oscillation region due to the attenuation factor and small group velocity.

\subsection{Anderson localization}
\begin{figure}[b]
	\includegraphics{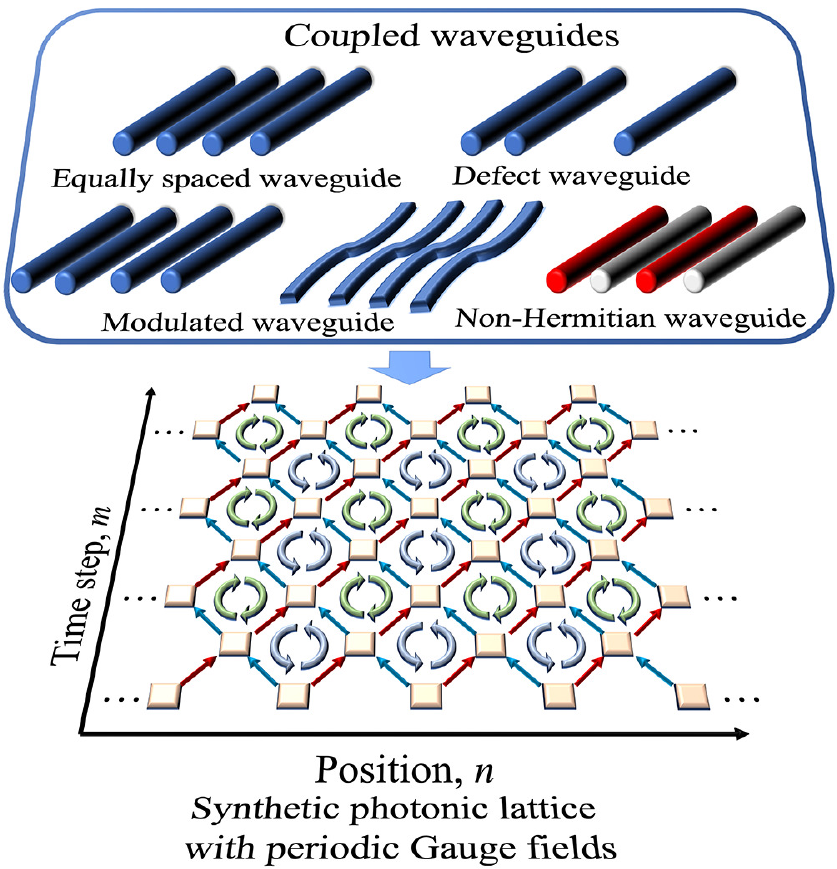}
	\caption{\label{perspect} Prospective map on the potential researches of the SPL with periodic gauge fields.}
\end{figure}
In the framework of coupled waveguides, Anderson localization can occurs for a disordered distribution of defects in potentials~\cite{segevnp13}. Naturally, a random modification in phase accumulations $2\phi$, Anderson localization can be realized in the SPL. The corresponding expression is
\begin{equation}
\psi_{m,n}=\left\{
\begin{array}{lr}
\phi-\delta\phi_n, & \mod(n,2)=0 \\
0, & \mbox{otherwise}
\end{array}
\right.
,\varphi_{n}=-\psi_{n},
\end{equation}
where $\delta\phi_n$ is a random number in the range ($-\Delta\phi/2,\Delta\phi/2$) over position site $n$.  The optical dynamics of point source are numerically simulated in such a SPL, as shown in Fig.~\ref{anderson}. One can see that the optical localization begins to emerge when the phase accumulations are weakly modulated. Besides, the majority of the light energy distributes in the position sites with gauge fields. For a increasing strength of maximum phase disorder $\Delta\phi$, the expanded region of light becomes narrow and the maximum intensity of light distributes around the injected site, manifesting the enhancement of localization strength. 

To quantify the strength of localization, it is customary to analyze the participation ratio $P_m=(\Sigma_n|v_n^m|^2)^2/\Sigma_n|v_n^m|^4$~\cite{derevyankosr18}. The larger $P_m$ indicates the broader distribution of wave packet. Figure~\ref{anderson}(e) plots the calculated $P_m$ as a function of $m$ for different maximum phase disorders corresponding to Figs.~\ref{anderson}(a-d). The values of $P_m$ enlarge at the beginning and finally reach a relative steady states. Consistently with the numerical results, with a larger value of $\Delta\phi$, the smaller participation ratio $P_m$ is obtained.

\section{Discussion and Prospect}
According to the above discussion, the prime advantage of our SPL is the ability to directly visualize the highly resembled optical behaviors in the homogenous or modulated waveguide arrays. Basically, with position-relevant periodic gauge field, all the optical propagations in SPL obey the modulation regulations of the waveguide arrays. Therefore, we contribute that a connection between the SPL and one-dimensional waveguides array is established by the introduced periodic gauge field [see Fig.~\ref{perspect}]. As a translational symmetry, a large variety of optical phenomena in different typles of coupled waveguides can be realized with properly controlling the electronic signals imported into the modulators and coefficient of the coupler in a visible vision. For example, the controllable parameters enable the SPL to simulate optical phenomena in various coupled homogenous waveguides, defect waveguides and modulated waveguides flexibility. If the nonlinearity is considered, solitons in nonlinear waveguides can also be simulated with dispersion management. Besides, SPLs can be used to study non-Hermitian even ${\cal PT}$-symmetric waveguide arrays with proper configuration of modulated signals once the gain and loss can be easily induced by adding intensity modulators in the fiber loops. Further, SPL also provides a possibility to achieve the defect modes, surface modes or topological edge state when a position-relevant boundary of gauge fields is introduced. In addition, the modulation protocols of coupled waveguides enrich the styles of temproal pulse modulation via SPL. Last but not the least, our SPL paves a new way to study the impulse dynamics through the analogue of periodically modulated gauge fields in synthetic mesh lattice and direct photonic lattice.

\section{Conclusion}
We constructed a novel type of synthetic photonic lattice consisting of a position-dependently periodic gauge field \raisebox{0.3mm}{---}  two opposite phase accumulations arranged alternatively along the position axis \raisebox{0.3mm}{---} by tuning the phase modulators in the two fiber loops, in which the band structures and light behaviors are highly similar to those in an evanescently coupled one-dimensional waveguide array. By tuning the phase accumulations $2\phi$ and the coupling coefficient $\theta$ of the coupler, the band structure, translation and gap of the two bands in the first Brillouin zone can be altered flexibly. When the values of $\phi$ and $\theta$ are properly chosen, similar to discrete diffraction in direct one-dimensional waveguide arrays, light hops in the position sites with identical phase accumulations and nearly zero energy exists in other sites.  Besides, tailored Gaussian wave packet responses and Talbot effect are achieved in the SPL, which follow the same principle associated with coupled waveguides. Further study shows the Bloch oscillations and Anderson localizations can also easily be obtained in the linearly and randomly modified phase circumfluence based on periodic gauge field. Our proposed modulation method on the SPL provides a transplantable routine from one-dimensional coupled waveguides to SPL platform through transitional symmetry. Moreover, the proposed research with respect to waveguide arrays provides abundant modulation fashions that can be utilized in SPL for beam shaping or controlling in time domain.

\begin{acknowledgments}
This work is supported by National Nature Science Foundation of China (61905193); National Key R\&D Program of China (2017YFB0405102); Key Laboratory of Photoelectron of Education Committee Shaanxi Province of China (18JS113); Open Research Fund of State Key Laboratory of Transient Optics and Photonics (SKLST201805); Northwest University Innovation Fund for Postgraduate Student (YZZ17099). Xinyuan Qi acknowledge financial support from the China Scholarship Council.
\end{acknowledgments}

\nocite{*}

\bibliography{splrefs}

\end{document}